\documentclass[aps,showpacs,pre,superscriptaddress]{revtex4}%
\usepackage{amsfonts}
\usepackage{amsmath}
\usepackage{mathrsfs}
\usepackage{amssymb}
\usepackage{wasysym}
\usepackage{graphicx}%
\providecommand{\U}[1]{\protect\rule{.1in}{.1in}}

\begin{document}
\title{Quantum phase transitions of ultra-cold Bose system in non-rectangular optical
lattices }
\author{Lin Zhi}
\affiliation{Department of Physics, Shanghai University, Shanghai 200444, P.R. China}
\author{Jun Zhang}
\affiliation{Department of Physics, Shanghai University, Shanghai 200444, P.R. China}
\author{Ying Jiang}
\thanks{Corresponding author}
\email{yjiang@shu.edu.cn}
\affiliation{Department of Physics, Shanghai University, Shanghai 200444, P.R. China}
\affiliation{Key Lab for Astrophysics, Shanghai 200234, P.R. China}

\begin{abstract}
In this paper, we investigate systematically the
Mott-insulator-Superfluid quantum phase transitions for ultracold
scalar bosons in triangular, hexagonal, as well as Kagom\'e
optical lattices. With the help of field-theoretical effective
potential, by treating the hopping term in Bose-Hubbard model as
perturbation, we calculate the phase boundaries analytically for
different integer filling factors. Our analytical results are in
good agreement with recent numerical results.

\end{abstract}


\maketitle

\section{Introduction}

The physics of dilute ultracold quantum gases in optical lattices
has grown to one of the most fascinating fields over the last
decade \cite{bloch-rmp2008,jaksch-zoller-ap,lewenstein}. In the
case of bosons, the delicate balance between the atom-atom on-site
interaction and the hopping amplitude leads to a quantum phase
transition \cite{sachdev-qptbook} between two distinct phases
\cite{fisher,zoller,greiner}. When the on-site interaction is
small compared to the hopping amplitude, the ground state is
superfluid, as the bosons are phase coherent and delocalized. In
the opposite limit of the strong on-site repulsion, the ground
state is a Mott-insulator, as every boson is trapped in one of the
potential minima. Meanwhile, this clean defectless setup, which
allows for precise control of its paramters, has opened up testing
ground for quantum many-body physics, i.e. the so-called quantum
simulation \cite{lewenstein}.

While most experiments with ultracold atoms up to date have been
performed in simple cubic lattices due to the ease of their
experimental implementation \cite{petsas-pra1994}, recent work
have explored ultra-cold atoms in non-standard optical lattices
such as triangular\cite{sengstock-njp} and hexagonal optical
lattices \cite{sengstock-hexagonal}, even the Kagom\'e optical
lattices has also been recognized couple of months ago
\cite{kagome-lattice}. In fact, due to the complex of the lattice
structure in these systems, novel and rich new phases will be
exhibited. Hence, to determine the quantum phase diagrams
analytically in these systems becomed a major problem and need to
be investigated systematically.

Actually, there are two main analytical methods which can be used
to determine the phase boundaries of Bose gases. One is the
mean-field theory  \cite{fisher}, the other one is the so-called
strong-coupling expansion \cite{freericks-1}. However, comparison
with the Monte Carlo data \cite{capogrosso-1} shows that the
mean-field theory underestimates the location of the phase
boundary while the strong-coupling expansion goes in the opposite
direction. More recently, an alternative analytical treatment
based on the effective potential and Rayleigh-Schr\"odinger
perturbation theory has been presented \cite{axel-09}, this novel
method may, in principle, yields analytical results for phase
boundaries at arbitrary dimension and lobe number in arbitrarily
high order accuracy.

In this paper, with the help of this novel systematic approach, we
are going to determine the Mott-Insulator-Superfluid (MI-SF)
quantum phase boundaries of ultracold scalar Bose systems in
triangular lattice, hexagonal lattice and Kagom\'e lattice
analytically, and present the corresponding expressions of the
phase boundaries. Comparing to numerical solutions, the relative
deviation of our third-order analytical results is less than 10\%.

\section{The model and the effective potential method}

A system of spinless bosons trapped in a homogeneous optical
lattice can be described by the simple yet nontrivial Bose-Hubbard
Hamiltonian \cite{bloch-rmp2008,bloch} which reads
\begin{align}
H_{\mathrm{BH}} = -t\sum_{\langle i,j \rangle}\hat{a}_{i}^{\dagger}\hat{a}%
_{j}+\sum_{i}\frac{U}{2}\hat{n}_{i}(\hat{n}_{i}-1)-\mu\hat{n}_{i}
\label{bose-hubbard-hamiltonian}%
\end{align}
with $t$ being nearest neighbor hopping parameter while $U$
denoting the strength of on-site repulsion between two atoms,
$\hat{n}_{i}= \hat{a}_{i}^{\dagger}\hat{a}_{i}$ are the
corresponding particle number operator. $\mu$ is the chemical
potential. When the depth of the optical lattice wells is
increased, the hopping parameter $t$ decreases exponentially while
$U$ increases linearly \cite{zwerger-1}.

In order to investigate the superfluid-Mott insulator phase
transition systematically and to determine the corresponding phase
boundary analytically in a more accurate way, we are going to
tackle this issue via field theory, or namely, the effective
potential method \cite{axel-09}. To this end, we add for the
moment additional source terms with strength $J$ and $J^{*}$ into
the Bose-Hubbard Hamiltonian as following
\begin{equation}
\hat{H}_{BH}(J^{\ast},J)=-t\sum_{\langle i,j\rangle}\hat{a}_{i}^{\dagger}%
\hat{a}_{j}+ \sum_{i}(J^{\ast}\hat{a}_{i} + J \hat{a}_{i}^{\dagger} )+\hat
{H}_{0}, \label{hamiltonian-with-source}%
\end{equation}
and treat the hopping term and external source terms as
perturbations. Here
\begin{equation}
\hat{H}_{0}=\sum_{i}\frac{U}{2}\hat{n}_{i}(\hat{n}_{i}-1)-\mu\hat{n}_{i}%
\end{equation}
is the unperturbed part of the Hamiltonian.

It is quite straightforward that the grand-canonical free energy
can be presented as power series of both the hopping parameter $t$
and the source $J,J^{\ast}$ via Taylor's expansions. Since the
unperturbed ground states are Mott states with the same occupation
number on each site, i.e. the unperturbed state is local, $J$ and
$J^*$ can only appear in pair in the expansion. After regrouping
the terms in the free energy with respect of $J$ and $J^{*}$, the
free energy reads
\begin{equation}
F(J^{\ast},\,J,\,t)=N_{s}\left(\,F_{0}(t)+
\sum_{p=1}^{\infty}c_{2p}(t)\mid J
\mid^{2p}\,\right)\,, \label{free-energy}%
\end{equation}
with the expansion coefficients
\begin{equation}
c_{2p}(t)=\sum_{n=0}^{\infty}(-t)^{n} \,\alpha_{2p}^{(n)}
\label{expansion-coefficient-c2p}%
\end{equation}
being power series of hopping parameter $t$, $N_s$ is the total
number of lattice sites.

The superfluid order parameter $\psi=\langle\hat{a}_{i}\rangle$
and its complex conjugate can then be calculated by
\cite{H.kleinert 09,J.Zinn-Justin 09}
\begin{equation}
\psi=\frac{1}{N_{s}}\frac{\partial F(J^{\ast},\,J,\,t)}{\partial J^{\ast}},
\quad\psi^{\ast}=\frac{1}{N_{s}}\frac{\partial F(J^{\ast},\,J,\,t)}{\partial
J}. \label{psi-definition-j}%
\end{equation}

With the help of the above formula, a Legendre transformation of the
grand-canonical free energy can thus be conducted, leading to the effective
potential as follows:
\begin{equation}
\Gamma{(\psi,\,\psi^{\ast})}=\frac{F}{N_{s}} - \psi^{\ast} J - \psi J^{\ast}.
\label{effective-potential-definition}%
\end{equation}
Substituting Eq.(\ref{free-energy}) into
Eq.(\ref{psi-definition-j}), the relation between the order
parameter and the external source strength is obtained and can
then be used to eliminate $J$ and $J^{*}$ in Eq.
(\ref{effective-potential-definition}), giving
\begin{align}
\Gamma= F_{0}(t)-\frac{1}{ c_{2}(t)}\mid\psi\mid^{2} +\, \frac{c_{4}(t)}{
c_{2}(t)^{4}}\mid\psi\mid^{4}\,+\cdots.
\end{align}

Apparently  the effective potential takes exactly the form of
$\phi^{4}$ theory. Indeed, from
Eq.(\ref{effective-potential-definition}), it is immediately
recognized that the external sources can be expressed as
\begin{equation}
\frac{\partial\Gamma(\psi,\,\psi^{\ast})}{\partial\psi^{\ast}}=-J, \qquad
\frac{\partial\Gamma(\psi,\,\psi^{\ast})}{\partial\psi}=- J^{\ast}.
\end{equation}
Recalling that the original physical system what we are interested
in is a system with vanishing external sources, this equation
shows that the physical states correspond to the saddle points of
the effective potential. As is known, in $\phi^{4}$ theory the
phase transitions is marked by the sign change of the coefficient
of the quadratic term, indicating that for our system the phase
boundary can be found by letting $\frac{1}{ c_{2}(t)}=0$, the
solution of this equation will provide us the critical value of
$t$.

Actually, as shown in Eq.(\ref{expansion-coefficient-c2p}),
$c_{2}(t)$ is a power series of $t$, thus the radius of
convergence of $c_{2}(t)$ will reveal the location of the phase
boundary. Accordingly, the $n$-th order approximation of the phase
boundary is given by
\begin{equation}
t_{c}^{(n)}= -\frac{\alpha_{2}^{(n-1)}}{\alpha_{2}^{(n)}}.
\end{equation}
This equation shows that, in principle, the phase boundary of the
Bose-Hubbard system can be calculated out analytically with
arbitrarily high accuracy via the field theory method or effective
potential method. It should be pointed out that in the pioneer
work \cite{axel-09} in this field, the expression of the phase
boundary is different from what we got above. The reason is that
in that work, $\frac{1}{ c_{2}(t)}$ was inconveniently expanded to
a power series of $t$. However, this inconvenience would not
affect the credit of that elegant work.

\section{Arrow-line diagram representation}

From the above discussion, we have translated the problem of
finding the phase boundary to the problem of calculating the
coefficient $\alpha_{2}^{(n)}$. These quantities can be calculated
via linked cluster expansion \cite{metzner} and
Rayleigh-Schr\"odinger perturbation theory, or namely the Kato's
formulation \cite{kato}.

Since the unperturbed state corresponding to the unperturbed
Hamiltonian $H_0$ is Mott state with all lattice sites occupied by
the same number of bosons, thus, according to Eq.
(\ref{hamiltonian-with-source}), in the free energy $F(J^*,J,t)$
 (\ref{free-energy}), between the bra and ket of the
unperturbed ground state, every non-zero contribution to
$\alpha_2^{(n)}$ includes exactly one creation operator at $i$-th
site (associated with $J_i$) and one annihilation operator at
$j$-th site (associated with $J^*_i$) as well as $n$ nearest
neighboring hopping processes (associated with $t$) connecting $i$
and $j$ sites via other neighboring lattice sites. $i$-th and
$j$-th sites could be the same.

\begin{table}[h!]
\caption{Diagrammatic expressions of $\alpha_{2}^{(n)}$ for
triangular
lattice, hexagonal lattice and Kagom\'{e} lattice}%
\begin{tabular}
[c]{|c|c|c|c|}\hline & triangular lattice & hexagonal lattice &
kagome lattice\\\hline
&  &  & \\
$\alpha_{2}^{(0)}$ & $\includegraphics[width=1.5cm]{1.eps}$ &
$\includegraphics[width=1.5cm]{1.eps}$ &
$\includegraphics[width=1.5cm]{1.eps}$\\\hline
&  &  & \\
$\alpha_{2}^{(1)}$ & $6\includegraphics[width=2cm]{2.eps}$ &
$3\includegraphics[width=2cm]{2.eps}$ & $4\includegraphics[width=2cm]{2.eps}$%
\\\hline
&  &  & \\
$\alpha_{2}^{(2)}$ &
$6\includegraphics[width=2cm]{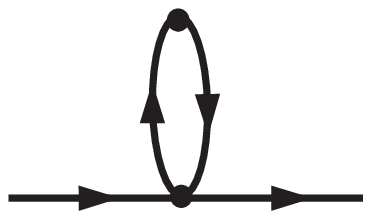}+30\includegraphics[width=2cm]{4.eps}$
&
$3\includegraphics[width=2cm]{3.eps}+6\includegraphics[width=2cm]{4.eps}$
&
$4\includegraphics[width=2cm]{3.eps}+12\includegraphics[width=2cm]{4.eps}$%
\\\hline
&  &  & \\
& $138\includegraphics[width=3cm]{5.eps}$ &
$12\includegraphics[width=3cm]{5.eps}$ &
$32\includegraphics[width=3cm]{5.eps}$\\
$\alpha_{2}^{(3)}$ &
$+30\includegraphics[width=2cm]{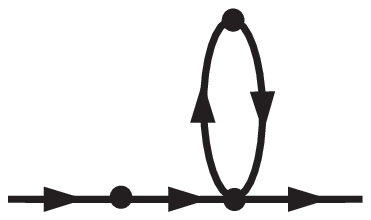}+30\includegraphics[width=2cm]{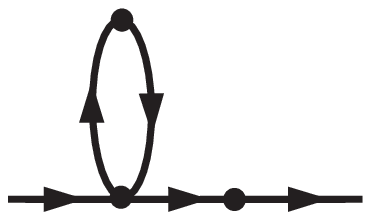}$
&
$+6\includegraphics[width=2cm]{7.eps}+6\includegraphics[width=2cm]{6.eps}$
&
$+12\includegraphics[width=2cm]{7.eps}+12\includegraphics[width=2cm]{6.eps}$\\
&
$+6\raisebox{-0.2cm}{\includegraphics[width=2cm]{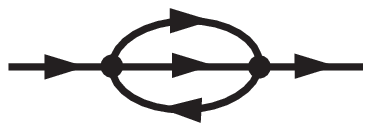}}+12\includegraphics[width=1.5cm]{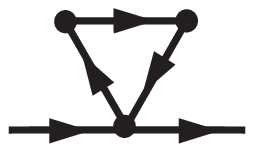}$
& $+3\raisebox{-0.2cm}{\includegraphics[width=2cm]{8.eps}}$ &
$+4\raisebox{-0.2cm}{\includegraphics[width=2cm]{8.eps}}+4\includegraphics[width=1.5cm]{9.eps}$%
\\\hline
\end{tabular}
\label{table-1}
\end{table}

When denoting a creation operator on a site by an arrow line
pointing in this site and an annihilation operator by an arrow
line pointing out of this site, each contribution of
$\alpha_2^{(n)}$ can be sketched as an arrow-line diagram composed
of $n$ oriented internal lines connecting the vertices and two
external arrow lines, the vertices in the diagram correspond to
respective lattice sites, oriented internal lines stand for the
hopping process between sites, the two external arrow lines are
directed into and out of a lattice site, respectively,
representing creation and annihilation operators associated with
the external sources. Each vertex in a diagram must has equal
number of pointing-in lines and pointing-out lines. Actually, the
diagrams are embedded in the lattices and, therefore, their
topologies are determined by the topologies of underlying
lattices. Table \ref{table-1} presents all diagrams as well as the
associated prefactors for the first four orders of
$\alpha_2^{(n)}$ in triangular, hexagonal, and Kagom\'e lattices.

In order to determine the MI-SF quantum phase transitions of
scalar Bose system in these non-rectangular lattices, the
corresponding $\alpha_2^{(n)}$ in Table I need to be calculated
explicitly, namely, these arrow-line diagrams should be translated
to proper perturbation formulae.

\section{Line-dot diagrams of perturbation calculation}

Actually, perturbation calculation can be endowed with a line-dot
diagrammatic representation \cite{axel-09} which can be used to
simplify the calculation of above mentioned $\alpha_{2}^{(n)}$
diagrams. In order to make the paper self-contained, let us here
give a brief review of this graphic representation of the
perturbation theory, for detail technique please refer to
Ref.\cite{axel-09}.

Suppose the total Hamiltonian is $H=H_{0}+gV$ in which $V$ is a
perturbation with small parameter $g$. In general, the eigenenergy
of the Hamiltonian $H$ is
$E_{n}=\sum_{i=0}^{\infty}g^{i}E_{n}^{\left(  i\right)
}$ , where%
\begin{equation}
E_{n}^{\left(  i\right)  }=\left\langle \Psi_{n}^{\left(  0\right)
}\right\vert V\left\vert \Psi_{n}^{\left(  i-1\right)
}\right\rangle \label{eigenvalue-i}
\end{equation}
with%
\[
\left\vert \Psi_{n}^{\left(  i\right)  }\right\rangle =\sum_{m\neq
n}\left\vert \Psi_{m}^{\left(  0\right)  }\right\rangle \frac{\left\langle
\Psi_{m}^{\left(  0\right)  }\right\vert V\left\vert \Psi_{n}^{\left(
i-1\right)  }\right\rangle }{E_{n}^{\left(  0\right)  }-E_{m}^{\left(
0\right)  }}-\sum_{j=1}^{i}E_{n}^{\left(  j\right)  }\sum_{m\neq n}\left\vert
\Psi_{m}^{\left(  0\right)  }\right\rangle \frac{\left\langle \Psi
_{m}^{\left(  0\right)  }|\Psi_{n}^{\left(  i-j\right)  }\right\rangle }%
{E_{n}^{\left(  0\right)  }-E_{m}^{\left(  0\right)  }},
\]
where $\left\vert \Psi_{n}^{\left(  0\right)  }\right\rangle $ is
the $n$-th eigenstate of unperturbed part $H_{0}$ while
$E_{n}^{\left(  0\right)  }$ is the corresponding eigenenergy.
These complicated forumulae have a simple brief graphic
representation. The detailed expression of $E_{n}^{\left( i\right)
}$ in Eq.(\ref{eigenvalue-i}) can be represented as a combination
of dots and lines by making use of the following rules: (1) each
dot represents an interaction $V$; (2) $p$ internal lines
connecting two adjacent dots stand for $\sum_{m\neq
n}\frac{1}{\left( E_{n}^{\left( 0\right) }-E_{m}^{\left( 0\right)
}\right) ^{p}}\left\vert \Psi_{m}^{\left( 0\right) }\right\rangle
\left\langle \Psi_{m}^{\left(  0\right) }\right\vert $; (3)
$\left\langle \Psi_{n}^{\left(  0\right) }\right\vert $ and
$\left\vert \Psi_{n}^{\left( 0\right) }\right\rangle $ can only be
denoted by the left-external and right-external lines,
respectively. From the third rule, it is easy to recognize that in
the diagrammatic representation of $E_{n}^{\left(  i\right)  }$
there are some graphs which consist of disconnected parts, the
sign in front of this type of graphs is $\left(  -1\right)
^{s-1}$, where $s$ is the number of the disconnected parts in this
graph.

To illustrate these rules more clearly, as an example, let us
concentrate on $E_{n}^{\left(  3\right)  }$. From Eq.
(\ref{eigenvalue-i}), it reads%
\begin{align*}
E_{n}^{\left(  3\right)  }  &  =\sum_{m_{1}\neq n}\sum_{m_{2}\neq
n}\left\langle \Psi_{n}^{\left(  0\right)  }\right\vert V\frac{\left\vert
\Psi_{m_{2}}^{\left(  0\right)  }\right\rangle \left\langle \Psi_{m_{2}%
}^{\left(  0\right)  }\right\vert }{E_{n}^{\left(  0\right)  }-E_{m_{2}%
}^{\left(  0\right)  }}V\frac{\left\vert \Psi_{m_{1}}^{\left(  0\right)
}\right\rangle \left\langle \Psi_{m_{1}}^{\left(  0\right)  }\right\vert
}{E_{n}^{\left(  0\right)  }-E_{m_{1}}^{\left(  0\right)  }}V\left\vert
\Psi_{n}^{\left(  0\right)  }\right\rangle \\
&  -\sum_{m\neq n}\left\langle \Psi_{n}^{\left(  0\right)  }\right\vert
V\left\vert \Psi_{n}^{\left(  0\right)  }\right\rangle \left\langle \Psi
_{n}^{\left(  0\right)  }\right\vert V\frac{\left\vert \Psi_{m}^{\left(
0\right)  }\right\rangle \left\langle \Psi_{m}^{\left(  0\right)  }\right\vert
}{\left(  E_{n}^{\left(  0\right)  }-E_{m}^{\left(  0\right)  }\right)  ^{2}%
}V\left\vert \Psi_{n}^{\left(  0\right)  }\right\rangle,
\end{align*}
then, by utilizing the rules mentioned above, we have%
\begin{equation}
E_{n}^{\left(  3\right)  }=\;\;\includegraphics[width=2cm]{d2.eps} -
\raisebox{-0.3cm}{\includegraphics[width=1.5cm]{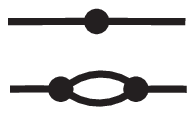}}. \label{en3}%
\end{equation}

This procedure can be extended to cases with several perturbation
terms \cite{axel-09}. However, in those cases, dots may represents
different perturbations and should be labelled by different
numbers, hence the line-dot diagrams mentioned above are now
extended to sums of all possible permutations with respective
topology. All the diagrams must be read from right to left to
present correct formula.

\section{Quantum Phase diagrams of Bose systems in non-rectangular optical lattices}

With this line-dot diagrammatic representation of perturbation
theory in hand, we now turn to our problem to see how the arrow
diagrams included in $\alpha _{2}^{\left( n\right) }$ can be
translated into the above mentioned line-dot diagrams.

As is known, each $\alpha _{2}^{\left( n\right) }$ consists of
exactly one creation operator (associated with $J$), one
annihilation operator (associated with $J^*$), and $n$ hopping
operators (associated with $t^n$). Moreover, since the unperturbed
ground state is local states, thus all these terms should be
treated as different perturbations, i.e., each arrow in the arrow
diagram of $\alpha _{2}^{\left( n\right)  }$ corresponds to a dot
in the respective line-dot diagram. Then with these dots, for each
arrow-line diagram we draw all possible topologically different
line-dot diagrams. The sum of these diagrams gives the
corresponding result.  To illustrate this more clearly, let us
take
\includegraphics[width=0.7cm]{9.eps} as an example. This arrow
diagram is a part of  $\alpha _{2}^{\left(  3\right)  }$ in
triangular lattice and Kagom\'e lattice. As discussed above, we
label the five arrow lines explicitly as
\begin{equation}
\includegraphics[width=2cm]{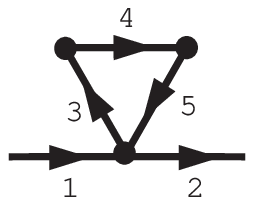}
\end{equation}
These five processes correspond to five distinguishable points in
line-dot diagrams. According to the translation rules discussed
above, this arrow diagram can be reformulated in terms of line-dot
diagrams as
\begin{eqnarray}
\raisebox{-0.5cm}{\includegraphics[width=2cm]{9-1.eps}}&=&\includegraphics[width=2cm]{d7.eps}
\;-\;\raisebox{-0.2cm}{\includegraphics[width=2cm]{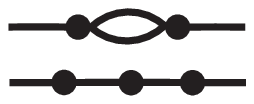}}\;-\;\raisebox{-0.2cm}{\includegraphics[width=2cm]{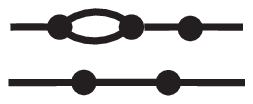}}\;-\;\raisebox{-0.2cm}{\includegraphics[width=2cm]{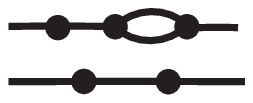}}.
\end{eqnarray}
Here, we should keep in mind that all these unnumbered line-dot
diagram in the equation stand for the sum of all possible
permutations. According to the line-dot diagram rule, all these
diagrams can be calculated. Since here we are only concentrated on
the properties of ground state, the calculation is relative
straightforward.

\begin{figure}[h!]
\includegraphics[width=0.5\linewidth]{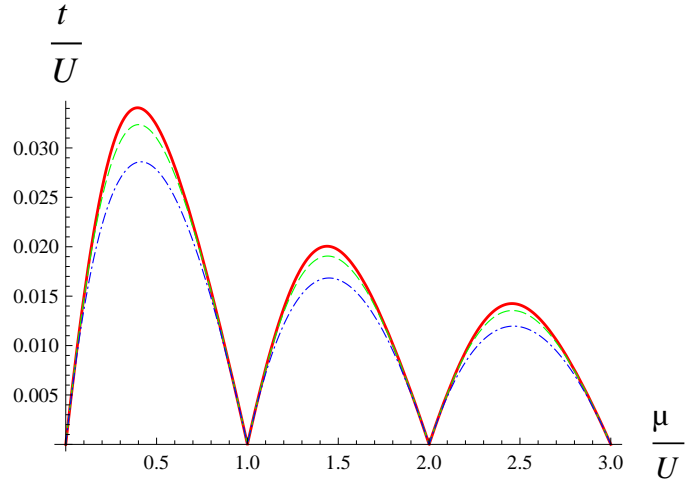}
\caption{(color online) The phase boundaries of MI-SF quantum
phase transition of Bose system in a triangle lattice at $T=0$.
The red solid line is the third-order calculation, the green
dashed line is the second-order
result, while the blue dot dashed line is the first-order (mean-field) result. }%
\label{triangular-phase-boundary}%
\end{figure}

By making use of the technique mentioned above, after a laborious
yet straightforward calculation, the third order approximation of
the phase boundaries in $t/U$-$\mu/U$ plane for scalar bosons in a
triangular lattice with different integer
filling factors are given by%
\begin{eqnarray}
t_{3} & = & -\frac{3(-3+\tilde{\mu})(-1+\tilde{\mu})\tilde{\mu}(-3-5\tilde{\mu}-7\tilde{\mu}^{2}-4\tilde{\mu}^{3}+3\tilde{\mu}^{4})U}%
{2(81+108\tilde{\mu}+135\tilde{\mu}^{2}+438\tilde{\mu}^{3}+164\tilde{\mu}^{4}-580\tilde{\mu}^{5}+266\tilde{\mu}^{6}-36\tilde{\mu}^{7})}%
,\ \ \ \mathrm{for\ \ }n=1 \nonumber \\
t_{3} &
=&\Big[(-4+\tilde{\mu})(-2+\tilde{\mu})(-1+\tilde{\mu})(1+\tilde{\mu})(12-129\tilde{\mu}+119\tilde{\mu}^{2}-37\tilde{\mu}^{3}
\nonumber \\
& &-20\tilde{\mu}^{4}+7\tilde{\mu}^{5})U\Big]\Big/\Big[2(-3056+13152\tilde{\mu}-23587\tilde{\mu}^{2}+7926\tilde{\mu}^{3} \nonumber \\
& &+13446\tilde{\mu}^{4}%
-11688\tilde{\mu}^{5}+806\tilde{\mu}^{6}+1836\tilde{\mu}^{7}-623\tilde{\mu}^{8}+60\tilde{\mu}^{9})\Big],\
\ \mathrm{for\ \ }n=2 \nonumber \\
t_{3}  &  =&\Big[ (-5+\tilde{\mu})(-3+\tilde{\mu})(-2+
\tilde{\mu})\tilde{\mu}(1080-1781\tilde{\mu}+776\tilde{\mu}^{2}
+14\tilde{\mu}^{3} \nonumber \\
& &-82\tilde{\mu}^{4}+13\tilde{\mu}^{5})U\Big]\Big/ \Big[ 2(-129600-72360\tilde{\mu}+707001\tilde{\mu}^{2}-859788\tilde{\mu}^{3}\nonumber \\
& & +389390\tilde{\mu}^{4}%
-13498\tilde{\mu}^{5}-51903\tilde{\mu}^{6}+19658\tilde{\mu}^{7}-2980\tilde{\mu}^{8}+168\tilde{\mu}^{9})\Big],\
\ \mathrm{for\ \ }n=3
\end{eqnarray}
where $\tilde{\mu}=\mu/U$. These analytical results are shown in
Fig. \ref{triangular-phase-boundary} together with the results for
first order and second order calculation.

Our third-order analytical result shows that the tip of the $n=1$
Mott lobe phase boundary is located at $t_c/U=0.03406$, it has
relative deviation of 9.7\% from the systematic strong coupling
expansion result \cite{elstner-monien} and 9.4\% from recent
numerical result \cite{Nikalas Teichmann1}.

\begin{figure}[h!]
\includegraphics[width=0.5\linewidth]{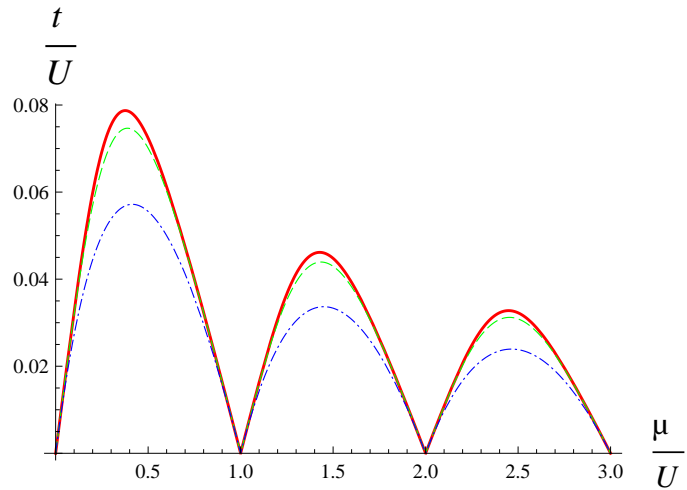}
\caption{(color online) The phase boundaries of MI-SF quantum
phase transition of Bose system in a hexagonal lattice at $T=0$.
The red solid line is the third-order calculation, the green
dashed line is the second-order
result, while the blue dot dashed line is the first-order (mean-field) result.  }%
\label{hexagonal-phase-boundary}
\end{figure}

The phase diagram of hexagonal lattice in  $t/U$-$\mu/U$ plane has
also been calculated out analytically via the same method, and the
results are presented in Fig. \ref{hexagonal-phase-boundary}. The
tip of the first Mott lobe phase boundary of hexagonal system is
located at $t_c/U=0.0787$, about 8.8\% relatively deviating from
the numerical result \cite{Nikalas Teichmann1}.

\begin{figure}[h!]
\includegraphics[width=0.5\linewidth]{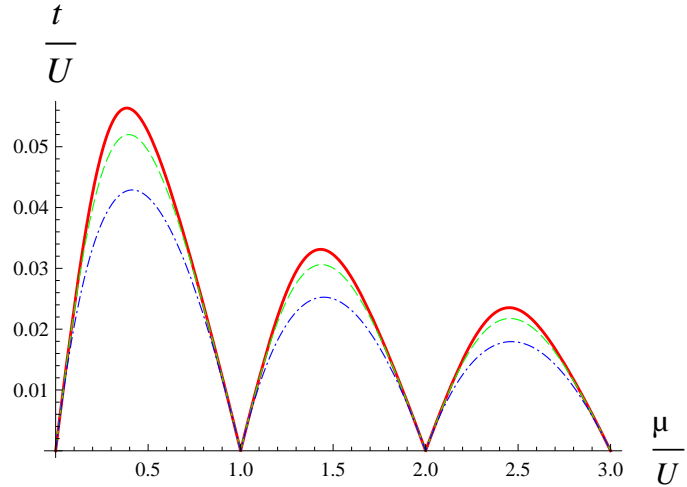}
\caption{(color online) The phase diagram of MI-SF quantum phase
transition of Bose system in a Kagom\'e lattice at $T=0$. The red
solid line is the third-order calculation, the green dashed line
is the second-order
result, while the blue dot dashed line is the first-order (mean-field) result.   }%
\label{kagome-phase-boundary}
\end{figure}

Kagom\'e-lattice many-body systems, due to their novelty, have
attracted much attention in condensed matter research in recent
decade. Naturally, it is worthy to spend more effort on quantum
simulating them via ultra-cold atoms in optical Kagom\'e lattices.
Fortunately, in recent days, this lattice has been realized in
experiment \cite{kagome-lattice}. Here, by making use the
field-theory method, we also calculate the quantum phase diagram
of scalar bosons in optical Kagom\'e lattice, the corresponding
analytical expressions of the phase boundaries for different Mott
lobes are
\begin{eqnarray}
t_c^{(3)}&=& -\frac{(-3+\tilde{\mu}) (-1+\tilde{\mu}) \tilde{\mu}
(-6-7 \tilde{\mu}-15 \tilde{\mu}^2-12 \tilde{\mu}^3+8
\tilde{\mu}^4) U}{72+24 \tilde{\mu}+113 \tilde{\mu}^2+482
\tilde{\mu}^3+203 \tilde{\mu}^4-620 \tilde{\mu}^5+274
\tilde{\mu}^6-36 \tilde{\mu}^7} \;\;\;\;{\rm for}\;\; n=1 ,
\end{eqnarray}
\begin{eqnarray}
t_c^{(3)}&=&\Big[ (-4+\tilde{\mu}) (-2+\tilde{\mu}) (-1+\tilde{\mu}) (1+\tilde{\mu})(40-372 \tilde{\mu}+377 \tilde{\mu}^2\nonumber\\
& & -101 \tilde{\mu}^3-60 \tilde{\mu}^4+20 \tilde{\mu}^5)\Big] U\Big/\Big(-10208+39280 \tilde{\mu}-69668 \tilde{\mu}^2+24200 \tilde{\mu}^3\nonumber\\
& & +41171 \tilde{\mu}^4-35302 \tilde{\mu}^5+1855
\tilde{\mu}^6+5788 \tilde{\mu}^7-1904 \tilde{\mu}^8+180
\tilde{\mu}^9\Big) \;\;\;\;{\rm for} \; \;n=2,
\end{eqnarray}
and
\begin{eqnarray}
t_c^{(3)}&=& \Big[(-5+\tilde{\mu}) (-3+\tilde{\mu}) (-2+\tilde{\mu}) \tilde{\mu} (1620-2669 \tilde{\mu}+1171 \tilde{\mu}^2+27 \tilde{\mu}^3\nonumber\\
& & -122 \tilde{\mu}^4+19 \tilde{\mu}^5)\Big] U \Big/\Big(-194400-100440 \tilde{\mu}+1027124 \tilde{\mu}^2-1245918 \tilde{\mu}^3\nonumber\\
& &+551310 \tilde{\mu}^4-4016 \tilde{\mu}^5-82749
\tilde{\mu}^6+30264 \tilde{\mu}^7-4519 \tilde{\mu}^8+252
\tilde{\mu}^9 \Big) \;\;\;\;{\rm for}\; \;n=3
\end{eqnarray}
respectively, where $\tilde{\mu}=\mu/U$.

\section{Summary}

In summary, with the help of effective potential theory, by
treating the hopping term in Bose-Hubbard model and external
sources as perturbations, and by conducting the linked cluster
perturbation calculation, we investigate systematically the
Mott-insulator-Superfluid quantum phase transitions for ultracold
scalar bosons in triangular and hexagonal lattices, the
corresponding analytical expressions of phase boundaries in these
systems and corresponding phase diagrams have been presented. By
comparing to recent numerical solutions, we have found that the
relative deviation of our third-order analytical results is less
than 10\%.

Stimulated by the recent experimental result \cite{kagome-lattice}
in which the Kagom\'e optical lattice was realized first time, we
have also calculated the MI-SF phase boundary for the scalar Bose
system in optical Kagom\'e lattice. Our result may serve as a
reference object for further studies. Meanwhile, as is known, in
the non-rectangular lattice systems, due to the nontrivial lattice
structures, novel and rich new phases will be exhibited,
especially in systems in triangular lattice and in Kagom\'e
lattice, since in these systems the geometrical frustration is
possible under some specific circumstance. Our result presented in
this paper may act as a starting point for further investigation
in this direction.

\section*{Acknowledgement}

Y.J. acknowledges A. Pelster and F. E. A. dos Santos for their stimulating
discussion. Work supported by Science \& Technology Committee of Shanghai
Municipality under Grant Nos. 08dj1400202, 09PJ1404700, and by NSFC under
Grant No. 10845002. Financial support from the Shanghai Leading Academic
Discipline Project( project Number: S30105) is also acknowledged.

\end{document}